\documentclass[10pt]{iopart}
\usepackage{iopams}
\usepackage{setstack}
\usepackage{graphicx}

\begin{document}
\title[Effective interaction potential and superfluidity of indirect excitons]{Effective interaction potential and superfluid-solid transition of spatially indirect excitons}
\author{A.~Filinov$^{1,2}$, P.~Ludwig$^1$, M.~Bonitz$^1$, and Yu.~E.~Lozovik$^2$}
\address{$^1$ Christian-Albrechts-Universit\"at zu Kiel, Institut
f\"ur Theoretische Physik und Astrophysik, Leibnizstrasse 15,
24098 Kiel, Germany}
\address{$^2$Institute of Spectroscopy RAS, Moscow region,
Troitsk, 142190, Russia}
\ead{filinov@theo-physik.uni-kiel.de}
\submitto{\JPA}
\begin{abstract}
Using an adiabatic approximation we derive an effective interaction potential for spatially indirect excitons in quantum well structures. Using this potential and path integral Monte Carlo simulations we study exciton crystallization and the quantum melting phase transition in a macroscopic system of 2D excitons. Furthermore, the superfluid fraction is calculated as a function of density and shown to vanish upon crystallization. We show that our results complement the previous studies of quantum dipole systems: We predict a second quantum phase transition -- melting of the crystal at high densities which originates from softening of the short-range part of the inter-exciton interaction.
\end{abstract}
\maketitle

\section{Introduction}
In recent decades systems of indirect excitons have been extensively studied experimentally~\cite{exp} due to the prospects to achieve superfluidity and Bose condensation.
Promising setups which have been successful in controlling the many-exciton state are electron-hole bilayers, e.g. ~\cite{senatore,njp} or single quantum wells
using the quantum Stark confinement (QSC). The latter will be studied in the present work.
As was shown in Ref.~\cite{Ludwig} trough the QSC one can simultaneously produce spatially indirect excitons and achieve their spatial localization. By varying the applied electric field one can control the exciton coupling parameter -- the ratio of the mutual exciton-exciton interaction to the confinement energy, as shown in our previous simulations for GaAs and ZnSe structures~\cite{Ludwig}.
With the increasing of inter-exciton coupling a striking phenomenon is expected -- spatial ordering of excitons into a crystalline lattice.
This quantum phase transition has been recently studied by a path integral Monte Carlo (PIMC) method for {\em trapped finite} Coulomb systems \cite{filinov_prl01}, systems of particles with dipole interaction~\cite{volk} and for the symmetric electron-hole bilayers~\cite{nelson,senatore,fil_bilayer,njp}.

In the case of interacting Bose particles, in addition, (partial) superfluidity is expected. Then during the crystallization transition one expects a gradual decrease of the superfluid fraction. This behaviour has recently been studied in detail by PIMC simulations for the Bose particles with the Coulomb interaction~\cite{fil08}. It was shown that the superfluid density can be concentrated either in the core or at the cluster boundary depending on the hexagonal order in the clusters which sensitively depends on the particle number.
Furthermore, a two-dimensional {\em homogeneous} dipole system of up to $400$ particles has recently been studied~\cite{Astrak,Buch}. The authors found a melting point
at a density of $n r_0^2=290 \pm 30$ or $D=\sqrt{n} r_0=17\pm 1$, for the definitions see Eq.~(\ref{r0}), and a step-like vanishing of the superfluid fraction at this point. In the present paper we test the validity of the dipole model for indirect excitons. In fact the effective exciton-exciton interaction $V_{xx}$ is a quantum-mechanical four-body   problem which has been studied e.g. by Zimmermann \cite{zimmermann} and may significantly deviate from a simple dipole form at distances comparable with the exciton dipole length. We therefore, devote special care to compute $V_{xx}$ from first principle PIMC simulations.
We find that this interaction allows for a second phase transition -- quantum melting of the exciton crystal by compression -- missing in dipole systems. Further, we found that for an effective exciton dipole length below $d \lesssim 6 a^*_B$ the crystalline phase vanishes.

Below we present results
for a model similar to the experimental setup of the Timofeev group~\cite{Timofeev}: a single ZnSe-based QW 
with indirect excitons produced by an electric field~\cite{Ludwig}  applied normal to a QW plane. In order to be able to perform simulations for a macroscopic ensemble of indirect excitons we apply a bosonic model. As shown by various groups~\cite{Shum,fil06}, in the moderate density regime the excitons are adequately treated as a composite particles obeying Bose statistics due to strong attractive interaction between electrons and holes. In the simulations of Ref.~\cite{fil06} we were able to test this approximation against the exact two-component fermion system in a broad range of densities.  In particular, we found that the bosonic model gives accurate predictions for the superfluid fraction once the excitons are in a strongly coupled -- low/moderate density regime. In contrast, in the weakly coupled -- high density regime close to or beyond the Mott density, the results are strongly affected by the Fermi statistics, and the bosonic approximation is no more valid. Thus, in the present analysis of moderate-density systems the bosonic model works well allowing us to study relatively large exciton ensembles without being hampered by the {\em fermion sign} problem.

\section{Model}
The general Hamiltonian for the system of $N_e$ electrons and $N_h$ holes ($N=N_e+N_h$) in the quantum well confinement and E-field can be written as
\begin{equation}
\hat H_{3D}=\hat H^{single}_{\parallel}+\hat H_z^{single}+W,
\label{H1}
\end{equation}
with the {\em single}  particle and {\em interaction} parts defined as
\begin{eqnarray}
&&\hat H^{single}_{\parallel}=\sum\limits_{i=1}^{N} \left[ -\frac{\hbar^2}{2 m^{\parallel}_{e(h)}}\nabla_{{\bf r}_i}^2 \right], \nonumber \\
&& \hat H_z^{single}=\sum\limits_{i=1}^{N} \left[ -\frac{\hbar^2}{2 m^{\perp}_{e(h)}}\nabla_{z_i}^2  + V^{QW}_{e(h)}(z_i)+V^{F}_{e(h)}\{E_z({\bf r}_i,z_i)\}   \right], \nonumber \\
&&W=\sum\limits_{i<j}^{N} V^{Coul}_{ij}, \quad V^{Coul}_{ij}= \frac{e_i e_j}{\epsilon} \left[(\mathbf r_i-\mathbf r_j)^2+(z_i-z_j)^2\right]^{-1/2}.
\end{eqnarray}
We consider a homogeneous electric field $E_z({\bf r}_i,z_i)= E_z(z_i)$,
vectors $\bf r$ denote 2D vectors in the QW plane, $V^{QW}$ is the QW confinement,  $V^{F}$ is the electrostatic potential energy due to the electric field and $\epsilon$ is the background dielectric constant.

To apply the approximation of bosonic excitons valid for low to moderate densities~\cite{fil06}, we want to reduce the 3D Hamiltonian~(\ref{H1}) to a 2D one, where all effects related to a particular width of the QW and the electric field strength will be combined in an effective inter-exciton interaction $V_{xx}(R)$. This becomes possible by using the adiabatic approximation~\cite{stark_eff,filinov_prb04}. This approach is justified for high values of the electric field $E_z$ which leads to a strong localization of electrons and holes at opposite edges of the QW. To be specific, the calculations below correspond to $L=30\ldots 120$nm wide ZnSe QWs and $E_z=20$kV.
Assuming the relation of energy scales, $\Delta \epsilon_i^{single} \gg U^{int}_{eh(ee,hh)}$, where
$\Delta\epsilon_{i}^{single}$ is the characteristic spacing of the quantized one-particle energy levels in $z$-direction und $U^{int}$ the interaction energy, we separate the out-of-plane motion and solve the 3D Bloch equation
for the $N$-particle density matrix $\rho_{3D}$
\begin{equation}
-\frac{\partial \hat \rho_{3D}(\beta)}{\partial \beta}=\hat H_{3D} \; \hat\rho_{3D}(\beta),
\label{bloch}
\end{equation}
in the adiabatic approximation, i.e.
\begin{equation}
\rho_{3D}(\beta)=\rho_{2D}(\mathbf r_1, \ldots,\mathbf r_{N},\beta) \, \prod\limits_{i=1}^{N_e} \rho_e(z_i,\beta)\, \prod\limits_{j=1}^{N_h} \rho_h(z_j,\beta),
\end{equation}
where $\beta =1/k_B T$ (below we drop the argument $\beta$). Now integrating out in Eq.~(\ref{bloch}) all $z$-dependencies, i.e. applying $\int \prod\limits_{i=1}^{N_e} dz_i \, \prod\limits_{j=1}^{N_h} dz_j$, we obtain a reduced $2D$ Bloch equation
\begin{equation}
-\frac{\partial \rho_{2D}}{\partial \beta}= \left(\sum\limits_{i=1}^{N} -\frac{\hbar^2}{2 m^{\parallel}_{e(h)}}\nabla_{{\bf r}_i}^2 + \sum\limits_{i<j}^N \tilde V_{\alpha_i\beta_j}(r_{ij}) + \sum\limits_{i=1}^N \epsilon_{i}^{single} \right) \rho_{2D},
\label{bloch2}
\end{equation}
where we introduced a smoothened Coulomb potential
\begin{equation}
\tilde V_{\alpha_i\beta_j}(r_{ij})= \int V^{Coul}_{ij} \, \rho_{\alpha_i}(z_i) \,
\rho_{\beta_j}(z_j) \, dz_i\, dz_j, \quad \alpha,\beta=e,h.
\label{smoothV}
\end{equation}
The densities $\rho_e(z_e)$ and $\rho_h(z_h)$ are found by solving a single-exciton problem in an electric field~\cite{stark_eff}. Also the exciton dipole moment, $\mu=e\cdot d$, follows directly from the electron and hole densities
\begin{equation}
d=\langle z_e\rangle - \langle z_h\rangle=\int z_e \rho_e(z_e) dz_e -\int z_h \rho_h(z_h) dz_h.
\label{dipole_m}
\end{equation}
The considered here low to moderate density regime leads also to another relation of energy scales, i.e. $E_B(X) \gg V_{xx}, k_B T$, where $E_B$ is the exciton binding energy. Under these conditions the excitons remain in their internal quantum states described by a two-body density matrix $\rho^{ex}(\mathbf r_{eh})$ throughout their interaction. This pair density matrix depends on the electron-hole separation $\mathbf r_{eh}=\mathbf r_{e}-\mathbf r_{h}$ and can be obtained numerically, e.g. with the matrix-squaring technique~\cite{Storer} applied to the interaction potential~(\ref{smoothV}). Using again the adiabatic approximation (now in the 2D plane) we write $\rho_{2D}$ as a product of a density matrix of an $N_x$ particle complex and relative density matrices $\rho^{ex}$ of $N_x$ excitons
\begin{equation}
\rho_{2D}=\rho_{2D}(\mathbf R^1, \ldots,\mathbf R^{N_x}) \, \prod\limits_{a=1}^{N_x} \rho^{ex}(r^a_{eh}).
\label{excH}
\end{equation}
Here we have assumed electrical neutrality, $N_e=N_h=N_x$, and introduced the  electron-hole pair coordinates related to the same exciton, $(\mathbf r_1,\ldots,\mathbf r_N)=\{(\mathbf r^a_e, \mathbf r^a_h)=(\mathbf R^a, \mathbf r^a_{eh})\}|_{a=1,\ldots,N_x}$, with the center of mass (c.o.m.) coordinates $\mathbf R^a=(m_e^{\parallel} \mathbf r^a_e +m_h^{\parallel} \mathbf r^a_h)/M_x, \; M_x=m_e^{\parallel}+m_h^{\parallel}$.
Certainly, the ansatz (\ref{excH}) implies that the excitons are stable against external perturbations and we are below the Mott density.

Now averaging Eq.~(\ref{bloch2}) over the relative degrees of freedom of excitons, i.e.
integrating over $\int \prod\limits_{a=1}^{N_x} d\mathbf r_{eh}^a$, we obtain the $N_x$-exciton Bloch equation depending on the c.o.m. coordinates
\begin{eqnarray}
&&-\frac{\partial \rho_{2D}(\mathbf R^1, \ldots,\mathbf R^{N_x})}{\partial \beta}= \left(\hat H^{eff} + E_x \right) \rho_{2D}(\mathbf R^1, \ldots,\mathbf R^{N_x}), \\
&&\hat H^{eff}=\sum\limits_{a=1}^{N_x} -\frac{\hbar^2}{2 M_{x}}\nabla_{{\bf R}^a}^2 + \sum\limits_{a<b}^{N_x} V_{xx}(R^{ab}), \label{compH} \\
&&E_x=\sum\limits_{a=1}^{N_x} \left\langle-\frac{\hbar^2}{2 \mu_{x}}\nabla_{{\bf r}_{eh}^a}^2 + \tilde V_{eh}(r^a_{eh})\right\rangle_{\rho^{ex}} + \sum\limits_{i=1}^N \epsilon_{i}^{single} .
\label{bloch3}
\end{eqnarray}
The interaction term in the {\em effective exciton} Hamiltonian $\hat H^{eff}$ is defined as the sum of the effective (adiabatically averaged) interactions of two electrons and two holes in excitons $a$ and $b$ ($a\neq b$)
\begin{equation}
V_{xx}(R^{ab})=\int \sum\limits_{\alpha,\beta=e,h}  \tilde V_{\alpha\beta}(|\mathbf r^a_{\alpha}-\mathbf r^b_{\beta} |) \, \rho^{ex}(r^a_{eh})\,  \rho^{ex}(r^b_{eh}) \, d\mathbf r^a_{eh}\, d\mathbf r^b_{eh}.
\label{trueU}
\end{equation}
The distances of two particles from different excitons can be expressed as
\begin{eqnarray}
&&\mathbf r_h^a-\mathbf r_h^b=\mathbf R^a - \mathbf R^b+m_e (\mathbf r_{eh}^a-\mathbf r_{eh}^b)/M_x,\nonumber \\
&&\mathbf r_e^a-\mathbf r_e^b=\mathbf R^a - \mathbf R^b-m_h (\mathbf r_{eh}^a-\mathbf r_{eh}^b)/M_x,\nonumber \\
&&\mathbf r_h^a-\mathbf r_e^b=\mathbf R^a - \mathbf R^b+(m_e \mathbf r_{eh}^a+m_h\mathbf r_{eh}^b)/M_x.
\end{eqnarray}
After integration, in Eq.~(\ref{trueU}) remains only the c.o.m. dependence on $R^{ab}=|\mathbf R^a - \mathbf R^b|$.

Thus, we have derived the effective Hamiltonian~(\ref{compH}) of composite particles. The interaction potential~(\ref{trueU}) generalizes the dipole potential used in the previous analysis of spatially indirect excitons~\cite{volk,Astrak,Buch}. The comparison of both is discussed below. The corresponding $N$-body problem~(\ref{compH}) can be solved with the path integral Monte Carlo technique which allows for a direct treatment of many-body correlation and bosonic exchange effects, for details see~\cite{cep,prokof,num_book}.

\section{Results}
We have performed PIMC simulations for a 2D homogeneous system with $N=60$ and $90$ bosonic excitons in a simulation box with periodic boundary conditions. The potential~(\ref{trueU}) has been divided into a short and long-range part,
$V_{xx}=(V_{xx}-V_D)+V_D$, with the dipole interaction, $V_D=(ed)^2/\epsilon r^3$, treated by the usual Ewald summation technique.
We used the following system of units: $r\rightarrow r/a_B^{*}$, $E\rightarrow E/Ha^*$, with the electron Bohr radius, $a_B^{*}=\hbar^2 \epsilon/m_e^{\parallel} e^2$, and the electron Hartree, $ Ha^{*}=e^2/\epsilon a_B^{*}$.  Here $M_x=m_e^{\parallel}+m_h^{\parallel}$ is the exciton mass. Parameters for typical semiconductor structures are listed in table~\ref{tab1}.
\begin{table}
\caption{\label{tab1}Semiconductor QW parameters. Masses are in units of the free electron mass $m_0$.}
\begin{indented}
\item[]\begin{tabular}{@{}lll}
\br
& GaAs/AlGaAs & ZnSe/ZnSSe \\
\mr
$\epsilon$ & 12.58 &   8.7\\
$m_e^{\parallel} $ & 0.0667 & 0.15  \\
$m_h^{\parallel} $ & 0.112 & 0.37 \\
$m_h^{\perp}$ & 0.377 & 0.86 \\
$M_x/m_e^{\parallel}$ & 2.68 & 3.46 \\
$a_B^*$ [nm] & 9.98 & 3.07  \\
$Ha^*$ [meV]& 11.47 & 53.93  \\
\br
\end{tabular}
\end{indented}
\end{table}
As it follows from the derivation in adiabatic approximation the Hamiltonian~(\ref{compH}) contains only the in-plane particles masses, $m^{\parallel}_{e(h)}$. The anisotropy of the parabolic bands both for electrons and holes, i.e. the out-of-plane effective masses, $m^{\perp}_{e(h)}$, are involved in the exciton solution in the $z$ direction. These masses determine the shape of the density matrices $\rho_e(z_e)$, $\rho_h(z_h)$ and hence indirectly influence the effective inter-exciton interaction via Eq.~(\ref{smoothV}). In particular, for a $20$kV/cm electric field applied to a $30$nm wide QW the calculations of Refs.~\cite{Ludwig,stark_eff} predict that the e-h separation~(\ref{dipole_m}) equals $d=15.78$nm for GaAs and $d=20.41$nm for ZnSe structures. Now comparing the ratio $\tilde d =d/a_B^*$ for both structures, we find $\tilde d=1.58$ and $\tilde d=6.65$, respectively. This shows that in ZnSe QW the excitons are more strongly coupled and it is easy to reach a crystalline regime as discussed below. Other advantages of using materials with larger effective masses are: a) increased stability of excitons due to higher binding energies and b) increased exciton life-time due to a better separation of carriers in the $z$-direction [the radiative life-time depends on the overlap of the density matrices $\rho_{e(h)}(z_{e(h)})$].  

In the following we, therefore, concentrate on the ZnSe structure. However, the results presented in the dimensionless units, $r/a_B^*$ and $E/Ha^*$, using table~\ref{tab1} can be applied to other materials as well. In particular,
the effective potential $V_{xx}$~(\ref{trueU}) already reduces to a dipole interaction $V_D=(ed)^2/\epsilon r^3$ at distances of about several exciton dipole moments $d$.
In this case the Hamiltonian~(\ref{compH}) can be brought to a universal dimensionless form using the scale~\cite{Buch}: $a_0=1/\sqrt{n}$, $E_0=\hbar^2 /M_x a_0^2$. Both systems of units are connected via relations
\begin{eqnarray}
&&n=\frac{1}{a_0^2}=\frac{1}{\pi \bar r^2}, \quad a_0=\sqrt{\pi} r_s \, a_B^*, \quad r_s=\bar r/a_B^*,  \nonumber \\
&&D=\frac{e^2 d^2}{\epsilon a_0^3\, E_0}=\left(\frac{M_x}{m_e^{\parallel}} \right) \tilde d^{\, 2} \frac{1}{\sqrt{\pi} \, r_s}, \quad \tilde d= d/a_B^*,\nonumber \\
&&\frac{E_0}{Ha^*}=\left(\frac{m_e^{\parallel}}{M_x}\right) \, \frac{1}{\pi r_s^2},
\label{r0}
\end{eqnarray}
where $n=N/(L_x L_y)$ is the number density.  

\begin{figure}[t]
\centering \includegraphics[width=0.70\textwidth]{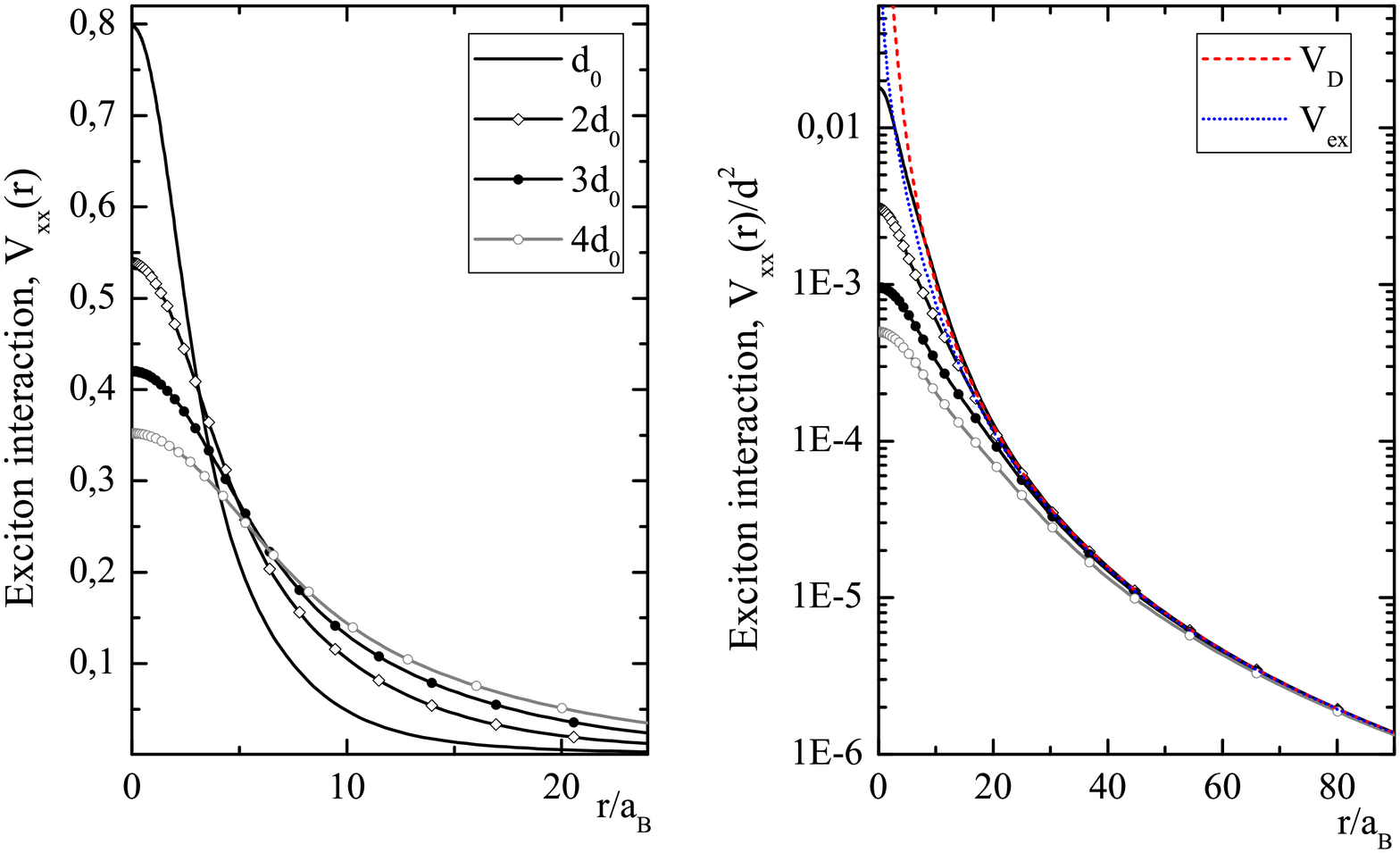}
\vspace{-0.5cm}
\caption{(Color online) Left: exciton interaction potential $V_{xx}(r)[Ha^*]$, Eq.~(\ref{trueU}), for several dipole moments $\tilde d$. Right:  $V_{xx}/\tilde d^2$ compared with the dipole potential, $V_D(r)=1/r^3$, and classical exciton potential, $V_{ex}=(2/r-2/\sqrt{r^2+\tilde d^2})/\tilde d^2$ (shown for $\tilde d=d_0$).}
\label{fex_int}
\end{figure}

In figure~\ref{fex_int} we show $V_{xx}(r) [Ha^*]$ for several e-h separations, $\tilde d= d_0, 2 d_0, 3 d_0, 4 d_0$ with $d_0=6.64848$ [$\tilde d=d_0$ corresponds to a $30$nm ZnSe QW and field strength $20$kV/cm]. In the right part we show, in addition, the dipole potential $V_D$ and the classical exciton pair potential 
(limit of ${\tilde d}\gg r$), $V_{ex}=2/r-2/\sqrt{r^2+\tilde d^2}$. While at large distances $V_{xx}$ agrees with $V_D$, for $r\lesssim 4{\tilde d}$, $V_{xx}$ is substantially weaker. 
Further, as one might expect from the dipole model, overall the interaction is stronger with increasing $d$. However, at small distances, $r  < 6 a_B^*$, $V_{xx}$  shows the opposite behavior which originates from the smoothening procedure~(\ref{trueU}) over the exciton relative density matrices: with increase of $\tilde d$ e-h pairs become more weakly bound, and the exciton in-plane size increases. This delocalization reduces the strength of the Coulomb interaction between two electrons and two holes, i.e. between two excitons. Vice versa, the stronger the binding of an e-h pair and its spatial localization the faster $V_{xx}$ approaches $V_{ex}$.

\begin{figure}[h]
\centering\includegraphics[width=0.4\textwidth]{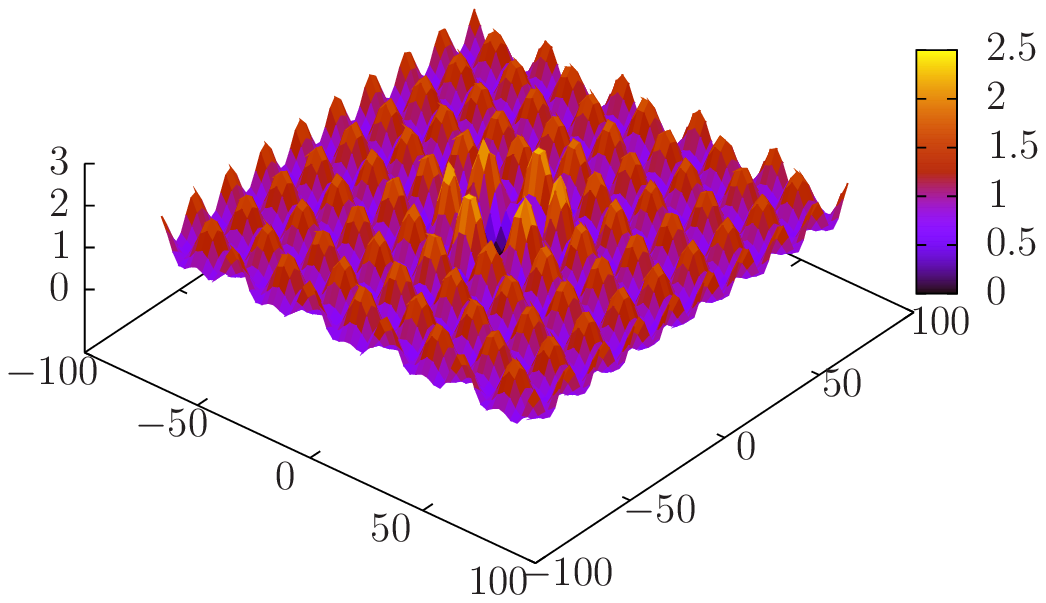}
 \includegraphics[width=0.31\textwidth]{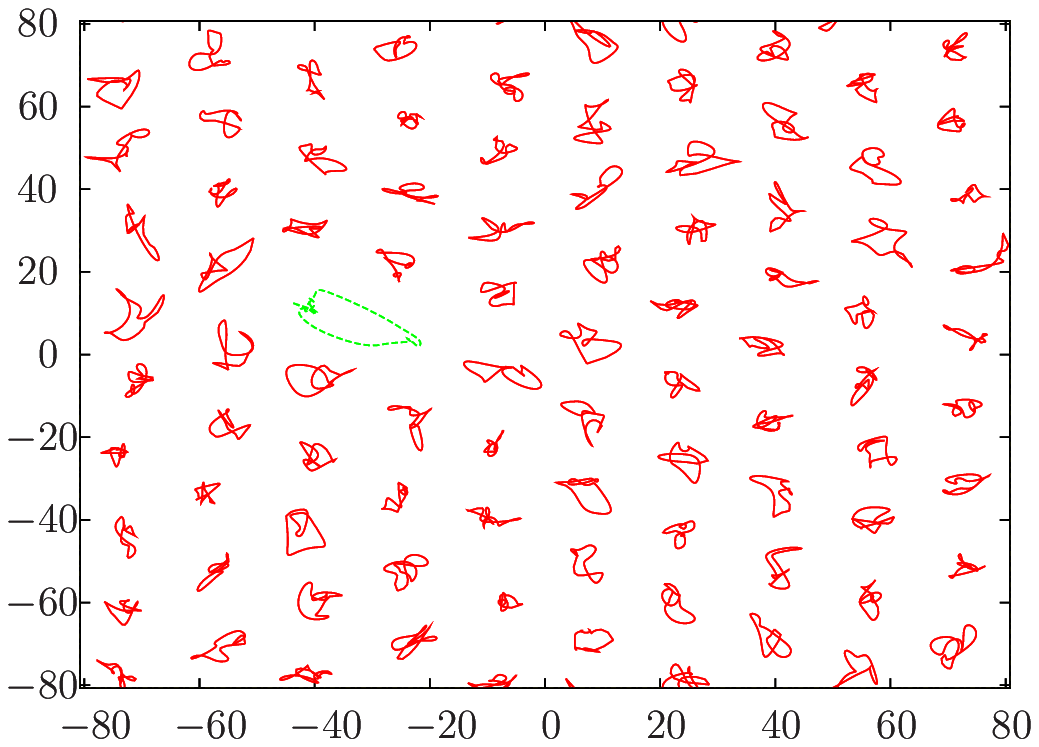}
 \includegraphics[width=0.4\textwidth]{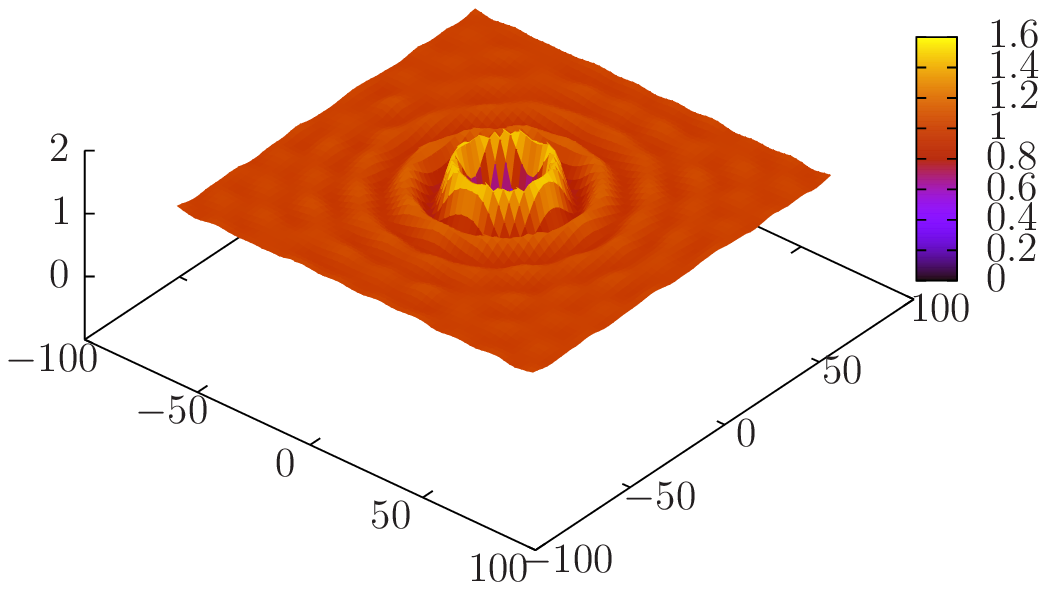}
 \includegraphics[width=0.31\textwidth]{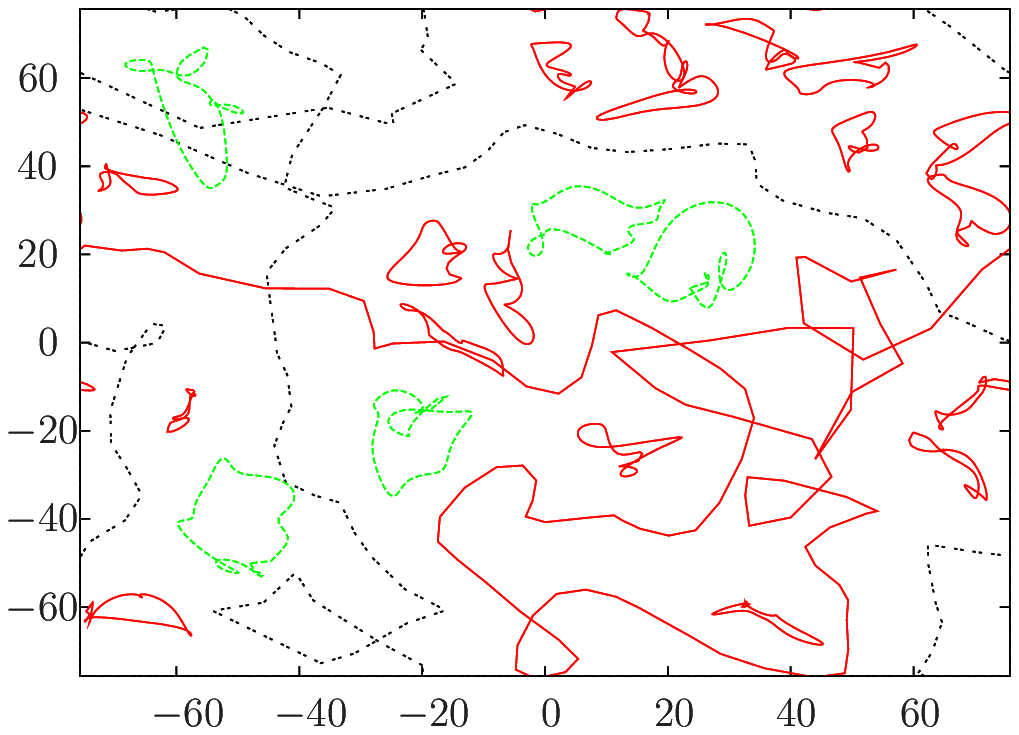}
\caption{(Color) Left: 2D pair distribution function at $r_s=10.0$ (top) in the solid and $r_s=9.5$ (bottom) in the superfluid gas phase. Right: snapshots from PIMC simulations. Trajectories of particles involved in permutations of different length are denoted with different colors. $\tilde d=3 d_0$}
\label{paird}
\end{figure}

\begin{figure}[th]
\centering\includegraphics[width=0.65\textwidth]{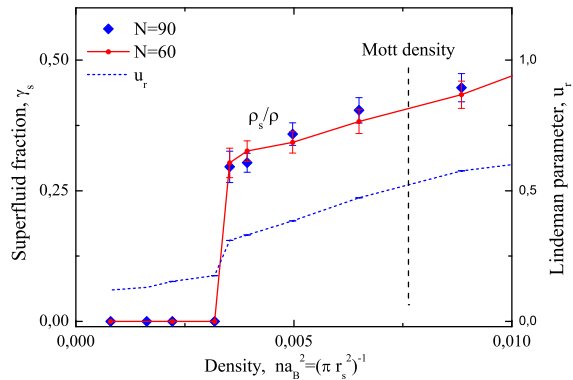}
\vspace{-0.5cm}
\caption{Superfluid fraction vs. density $n a_B^{*\, 2}$, for $N=60$ (circles) and $N=90$ (rombs) for $\tilde d=3 d_0$. Dotted line shows the Lindemann parameter $u_r$. Vertical line indicates the Mott density. The Bose model (superfluid data) is applicable only at lower densities \cite{fil06}.}
\label{superf}
\end{figure}

Let us now analyze the melting behavior of the exciton ensemble. In Refs.~\cite{Astrak,Buch}, crystallization of dipoles was observed at $D=17\pm 1$ which was identified by a break of the translational symmetry of the pair distribution functions (PDF) and the static structure factor. 
We now perform a similar analysis with the improved model interaction $V_{xx}(r)$, for $T=1/3000$Ha, and consider a density range of $\rho a_B^{*\, 2}=1/\pi r_s^2$, $r_s=5\ldots 12$. Simulations were performed for $\tilde d\,[d_0]=1,2,3,4$, and the results are shown in figures~\ref{paird} and \ref{superf}. At low densities where our potential is close to a dipole potential we observe similar results as in~\cite{Astrak,Buch}, i.e. crystallization of dipoles upon compression (not shown). But most importantly, at high density, we observe completely different behavior which is due to the weak potential at small $r$: the exciton crystal melts upon compression. 
This transition is clearly seen from the 2D PDF in figure~\ref{paird}(left) where an abrupt loss of a (quasi) long-range crystalline order is observed by a slight change in $r_s$ from $10$ (top) to $9.5$ (bottom). Simultaneously, in figure~\ref{paird}(right) we observe a topological change in the picture of the particle trajectories in the path integral representation \cite{cep}. While in the solid phase we observe only local exchanges of few particles, just after the melting transition the trajectories form macroscopically large permutation cycles crossing the edges of our periodic simulation cell. From the statistics of the flux of paths winding around the periodic cell one can estimate the fraction of the superfluid density~\cite{cep,poll}:
\begin{equation}
\gamma_s =\rho_s/\rho=M_x\langle W^2 \rangle/\hbar^2 \beta N_x, \quad \mathbf{W}=\sum\limits_{i=1}^{N_x} \int\limits_{0}^{\beta} dt \left[ d \mathbf{r}_i(t)/dt \right].
\end{equation}
Figure~\ref{superf} indicates a step-like increase of the superfluid density from zero up to about $35\%$ in the gas phase. The vertical dotted line shows the Mott density where the excitons pressure ionize and the bosonic model fails. The critical values $r^{h}_s$ for the exciton quantum melting transition at different $\tilde d$ are collected in table~\ref{tab2} together with the critical data $r_s^l$ for the dipole crystallization.
\begin{table}
\caption{\label{tab2}Interparticle distances at first $r_s^l$ and second $r_s^h$ phase  transition: superfluid gas -- exciton solid. $r_s^l$ are estimated from~(\ref{r0}) and $D \ge 17\pm 1$; $r_s^h$ are the PIMC results using $V_{xx}(r)$ in figure~\ref{fex_int}. The exciton solid exist for densities $n a_B^{*\, 2}=1/\pi r_s^2$, with $r_s^h \leq r_s \leq r_s^l$. Second colum, $L$, is the required ZnSe QW width at the field strength $E_z=20$kV/cm .}
\begin{indented}
\item[]\begin{tabular}{@{}llll}
\br
$\tilde d \, [d_0]$& $L$ [nm]& $r_s^l$ & $r_s^h$ \\
\mr
$1$ & 30&5.1(0.3) &  -- \\
$2$ & $\sim 50$ & 20.4 (1.2) & 10.0 (0.5)  \\
$3$ & $\sim 70$ & 45.90 (2.7) & 10.0 (0.5) \\
$4$ & $\sim 90$ & 81.6 (4.8) & 11.0 (0.5) \\
\br
\end{tabular}
\end{indented}
\end{table}
%

In conclusion, the derived exciton-exciton potential leads to completely different predictions for the phase diagram of bosonic excitons
compared to the dipole model. Due to the much softer Coulomb-like interaction at small distances, the exciton solid melts by compression, similar to a Wigner crystal of electrons \cite{filinov_prl01}.
Due to this fact it becomes possible to stabilize the exciton lattice only in a finite density interval [see table~\ref{tab2}]. Outside of this region the excitons exist in a superfluid gas phase.

Several heterostructures are candidates for the observed effect, but ZnSe is favorable due to its relatively high value of the dipole moment.
Using parameters from table~\ref{tab1} we estimate the exciton solid to exist in ZnSe (taking $\tilde d=2 d_0$) in a QW with
 $L \sim 50$nm between $0.81\leq \rho [10^{10}cm^{-2}] \leq 3.38$ at $T\lesssim$2K and in GaAs in a QW with $L\sim 148$nm  between  $0.77\leq \rho [10^9cm^{-2}] \leq 3.2$ and $T\lesssim 0.4$K.
%
While we have not considered disorder effects due to the imperfections of the QW planes they can be important. In our case of the electric field-induced indirect excitons electrons and holes are pushed to the QW edges and hence experience the influence of QW width fluctuations and impurities \cite{filinov_prb04}. As in the case of the electron Wigner crystal this can additionally stabilize the exciton solid at high densities.

\ack
Financial support by the Deutsche Forschungsgemeinschaft via SFB-TR24 grant A7 and FI 1252 is gratefully acknowledged.

\end{document}